\documentstyle[preprint,aps,prl]{revtex}

\begin{document}
\title{Phonon dispersion and electron-phonon coupling in MgB$_2$ and AlB$_2$}
\author{K.-P. Bohnen, R. Heid, and B. Renker}
\address{Forschungszentrum Karlsruhe, Institut f\"ur Ferstk\"orperphysik, 
P.O.B. 3640, D-76021 Karlsruhe, Germany}
\date{\today}

\maketitle
\begin{abstract}
We present a {\em first principles} investigation of the lattice
dynamics and electron-phonon coupling of the superconductor MgB$_2$ and
the isostructural AlB$_2$ within the framework of density
functional perturbation theory using a  mixed-basis pseudopotential
method.
Complete phonon dispersion curves and Eliashberg functions $\alpha^2$F
are calculated for both systems.
We also report on Raman measurements, which support the
theoretical findings.
The calculated generalized density-of-states for MgB$_2$ is in excellent
agreement with recent neutron-scattering experiments.
The main differences in the calculated phonon spectra and 
$\alpha^2$F are related to high frequency in-plane boron
vibrations. As compared to AlB$_2$, they are strongly softened in MgB$_2$
and exhibit an exceptionally strong coupling to electronic states at the
Fermi energy.
The total coupling constants are $\lambda_{MgB_2}=0.73$ and
$\lambda_{AlB_2}=0.43$. 
Implications for the superconducting transition temperature are briefly
discussed.
\end{abstract}

\newpage
\narrowtext





Recently superconductivity has been observed in MgB$_2$ 
with an exceptionally high transition temperature for such a simple
compound (T$_c\sim$39 K) \cite{Nagamatsu}. This system has in the
mean time received a lot of attention from many experimentalists
as well as theorists since understanding superconductivity in this
binary compound should be much easier than in the high-T$_c$
cuprate materials studied extensively for more than ten years now. This
is due to the simpler lattice structure and the missing complication
due to magnetism and strong electron-electron correlations.
Already early experiments
\cite{Budko,Finnemore,Rubio,Lorenz,Karapetrov,Kotegawa} as well as
theories \cite{Kortus,Belashchenko} have suggested that we are
dealing here with an s-wave superconductor based on strong
electron-phonon coupling although alternative coupling mechanisms
have been discussed too \cite{Hirsch,Imada}.

Modern band-structure calculations together with ab initio
determination of the phonon dispersion, electron-phonon coupling
and solution of the Eliashberg equations to calculate the
transition temperature should be the way to help understand the
role played by the electron-phonon coupling in this material.
Although during the past six weeks many contributions have been
published concerning the electronic bandstructure
\cite{Kortus,Belashchenko,An} only a few attempts have been made
to proceed along the above mentioned roadmap to a deeper
understanding of the electron-phonon coupling. Most
investigations are restricted to calculations of
the phonon modes for the $\Gamma$-point only \cite{Budko,Satta}.
Estimates of the coupling strength are therefore not
very accurate thus asking for a more complete treatment of the
electron-phonon coupling in the whole Brillouin zone (BZ).

We have carried out such a systematic study of the lattice dynamics and
the electron-phonon coupling using the mixed-basis pseudopotential
method. To get a better understanding of the relevance of the
electron-phonon mechanism we have studied two isostructural
systems: MgB$_2$, which is superconducting, and AlB$_2$, for which no
superconductivity has been found so far. This offers the possibility to
compare the phonon dispersion curves to identify those
which are strongly influenced by electron-phonon coupling even
without calculating the coupling strength. Furthermore, the
calculation of the Eliashberg function $\alpha^2$F for both
systems allows for a consistency check of the proposed phonon-mediated
pairing mechanism.

Only very recently after completion of our study we became aware
of a similar investigation by Y.~Kong et al.\ \cite{KongDolgov}
which calculated the full phonon dispersion, the electron-phonon
coupling and T$_c$ for MgB$_2$ within the LMTO-method. Our
calculations are very similar to those presented in
Ref.~\cite{KongDolgov},
however, due to the parallel treatment of MgB$_2$ and AlB$_2$
they offer additional information.

In contrast to all other calculations presented so far we have
structurally optimized both systems, thus we remain fully in the
framework set by the theory. For comparison and checks of
sensitivity of our results we have also studied certain phonon
modes using the experimental structure. All our calculations are
carried out using the mixed-basis pseudopotential method. For Mg
and Al we have used well tested pseudopotentials of
Martins-Troullier-\cite{Pelg} and BHS type \cite{Ho}. The boron
pseudopotential was constructed according to the Vanderbild
description \cite{Vanderbilt} which led to a fairly deep
p-potential for the boron atom which however could be dealt with
very efficiently due to the mixed-basis formulation.  The wave
functions were constructed from localized s and p functions at
the Boron sites supplemented by plane waves with an energy cut
off of 16 Ry. Detailed tests were carried out to assure
convergence with respect to the number of plane waves as well as
with respect to k-point sampling \cite{HeidBohnen}. Different
Monkhorst-Pack k-point sets have been used up to 13824 k-points
in the BZ together with a Gaussian smearing of 0.2 eV. The
calculation of the phonon dispersion is based on a recently
developed mixed-basis perturbation approach \cite{Heid} which also
allowed for the calculation of the electron-phonon coupling
\cite{Heidb}.
Finally, estimates of T$_c$ are obtained by solving the linearized form
of the Eliashberg equations \cite{Bergmann}.
Structural parameters were found to be fairly insensitive to the
k-point sets while individual phonon modes were very sensitive to
the sampling thus pointing to strong electron-phonon coupling.
All our calculations were carried out using the local
exchange-correlation potential of Hedin and Lundqvist \cite{Hedin}.

Since both systems crystallize in the so-called ''AlB$_2$
structure'' with alternating hexagonal (Mg,Al)-layers and
graphite-like B-layers structural optimization required
optimization of volume V and c/a ratio. Table 1 gives calculated
results for V and c/a as well as the bulk modulus B together with
the experimental values. For both systems a slightly smaller
volume has been found by the theoretical treatment compared to
the experimental one. This is a typical behavior in
LDA-calculations. The c/a ratio is slightly larger. The
electronic bandstructure obtained for MgB$_2$ is very similar to
the ones obtained by other groups \cite{Kortus,Belashchenko,Medvedeva}
with hole pockets around the $\Gamma$-point extending along the
$\Gamma$-A-direction. AlB$_2$ has a very similar bandstructure,
however, due to the additional electron density no hole pockets
around $\Gamma$ and the $\Gamma$-A-direction exist. 

In Table 2 we have summarized our phonon results for the $\Gamma$-point
obtained from calculations based on the experimental structure as well
as on the optimized
one. With the exception of the E$_{2g}$-mode (in-plane boron mode) all
other modes are fairly insensitive to structural changes as can be seen
from Table 2. The same holds also true for the sensitivity with respect
to k-point sets. All previous calculations of the $\Gamma$-point phonon
modes for MgB$_2$ agree very well with these results with the exception of
the E$_{2g}$-mode where shifts of $\pm$100 cm$^{-1}$ have been observed
\cite{Kortus,Satta,KongDolgov}. As extensive studies have shown \cite
{HeidBohnen} this is mostly an effect of k-point sampling and indicates
strong electron-phonon coupling for this mode. Comparing these results
with those obtained for AlB$_2$ the most dramatic changes occur again
for the E$_{2g}$ mode which stiffens substantially and becomes the
highest one, and the B$_{1g}$ mode (out of plane boron mode) which is
strongly softened.

We have measured the E$_{2g}$ mode by Raman scattering from samples of
commercially available MgB$_2$ and AlB$_2$ powders. 
The samples showed clear x-ray diffraction patterns and consisted of
crystalline grains up to 10 $\mu$m (MgB$_2$) and 50 $\mu$m (AlB$_2$).
Measured spectra showed only one prominent line in agreement with space
group P6/mmm and were not particularly polarization dependent.
Typical results
are presented in Fig.~1. The experimental result for AlB$_2$ is in
almost perfect agreement with the calculated one. For MgB$_2$ the Raman
line is very broad, however, the peak position agrees reasonably well with
the calculated one while the linewidth is likely to be due to strong
electron-phonon coupling (see discussion below).

Having studied carefully the $\Gamma$-point modes we calculated
also the full phonon dispersion for both systems. Results are
presented in Fig.~2 which were obtained by determining the
dynamical matrix on a (6,6,6) reciprocal lattice grid in the
hexagonal Brillouin zone. 
The results for MgB$_2$ agree very well with those of
Ref.~\cite{KongDolgov} and give thus additional credibility to the
theoretical treatment.

Comparing the phonon density-of-states shown in Fig. 2 certain
features are worth mentioning. The biggest difference between the
two systems is showing up in the intermediate region of phonon
frequencies where AlB$_2$ shows nearly a gap due to the strong
dispersion of the relevant modes while for MgB$_2$ this gap has
nearly completely disappeared. High density of states at the
upper and lower end of the frequency range show up in both
systems. To compare with recent measurements by inelastic neutron
scattering \cite{Osborn} we have calculated also the generalized
density-of-states (weighted by the inelastic scattering cross section
and the mass) broadened by a Gaussian with a width of 4 meV. The result
in Fig.~3
should be compared directly with Fig.~1 in Ref.~\cite{Osborn}. 
An almost perfect
agreement with respect to peak positions, shoulders and even with
the relative ratio of contributions is obvious.
We don't find any indication of a peak in the spectra near 17 meV, as
observed by Sato {\em et al.} \cite{Sato}.

In the dispersion curves shown in Fig.~2, the layered structure of the
crystals is reflected in a weak dispersion of the optical branches along
$\Gamma$-A and in an anisotropy in the slopes of the acoustic branches
in agreement with experiments \cite{Prassides}.
Besides many
similarities two very significant differences between AlB$_2$
and MgB$_2$ can be seen. The first one is related to the branches
which evolve from the doubly degenerate E$_{2g}$-mode at the
$\Gamma$-point. In MgB$_2$, these branches are
strongly renormalized towards lower frequencies mostly near $\Gamma$ and
along the $\Gamma$-A direction. This is probably related to the
hole pockets found in the electronic bandstructure of MgB$_2$
which are absent in AlB$_2$. The second distinctive difference is
the behavior of branches starting from the B$_{1g}$-mode which in
MgB$_2$ are significantly harder
in certain regions of the BZ than in AlB$_2$.

To address the superconducting properties we have calculated the so-called
Eliashberg function $\alpha^2$F($\omega$), using a very fine (36,36,36) 
k-point grid
in performing the Fermi-surface average of the electron-phonon matrix
elements \cite{Bauer}.
Results are presented in Fig. 4. For MgB$_2$ we
find indeed a very large contribution to $\alpha^2$F in the
intermediate region between 60 and 70 meV, which is mainly due to strongly
softened in-plane vibrations of the boron atoms as mentioned above.
At $\Gamma$, for the E$_{2g}$-mode this strong coupling results in a very
large linewidth of 15 meV in accordance with the broad feature seen in
the Raman spectrum (Fig.~1).
For AlB$_2$, in contrast, the biggest contributions
show up at very high frequencies as well as in the regime of the
acoustic modes, however, they are substantially smaller than the
main contribution in MgB$_2$. From the Eliashberg function we
calculate the electron-phonon coupling constant
\begin{displaymath}
\lambda=2\int^{\infty}_{o} d{\omega}
\frac{ \alpha^{2}F(\omega)}{\omega}
\end{displaymath}
which gives $\lambda_{MgB_2}$=0.73 and $\lambda_{AlB_2}$=0.43.
For the logarithmically averaged phonon frequencies as defined in
Ref.~\cite{KongDolgov} we find 60.9 meV and 49.9 meV for MgB$_2$ and
AlB$_2$, respectively.
The values for MgB$_2$ agree well with the results of
Ref.~\cite{KongDolgov}.
The fairly large value of $\lambda$ for AlB$_2$ however is unexpected.

Within the dirty limit of superconductivity \cite{Schrieffer}
the calculation of the transition temperature T$_c$ requires the
knowledge of $\alpha^2$F and the Coulomb pseudopotential $\mu^*$
\cite{Bergmann}.
As commonly done, we treat $\mu^*$ as an adjustable parameter.
A T$_c$ of 40 K for MgB$_2$ is obtained for $\mu^*$=0.05.
The same $\mu^*$-value leads to T$_c\sim$ 9 K for the
system AlB$_2$.
This result is in contrast to the experimental situation where no
superconductivity has been found so far for AlB$_2$. There are
basically two possibilities to reconcile the theoretical results
with the experimental situation. One possibility is a different
screening in AlB$_2$ compared to MgB$_2$ which has been
speculated about in Ref.~\cite{Voelker}, leading to different values for
$\mu^*$. Alternatively, the approximation of
an isotropic superconductor (dirty limit) might not hold. Due to
the fact that we have dealt with both systems on equal footing
new questions have emerged which need further studies.

We have presented here first principles calculations of the
phonon dispersion and electron-phonon coupling for two systems
MgB$_2$ and AlB$_2$ which crystallize in the same lattice
structure, however, which have fairly different phonon dispersion
curves.
These results are in excellent agreement with measured
quantities. The calculation of the superconducting temperature,
however, has a problem which might be due to approximations
involved or due to the fairly restricted knowledge about the
Coulomb pseudopotential $\mu^*$.

The authors would like to thank Dr. Reichardt, Dr. Pintschovius
and Dr. Schweiss for many helpful discussions.

\begin{table}[h]
   \caption{
 Structural parameters of the optimized geometries. Experimental values
(in brackets) are taken from Refs.~\protect\cite{Wyckoff,Prassides}.
}
\begin{tabular}{lccc}
  & V (\AA$^3$) & c/a & B (Mbar) \\
\tableline
MgB$_2$ & 28.481 (29.010) & 1.153 (1.142) & 1.47 (1.20) \\
AlB$_2$ & 24.617 (25.578) & 1.09 (1.084) & 1.84        
\end{tabular}
\end{table}  
\begin{table}[h]
   \caption{
Comparison of the calculated $\Gamma$-point phonon frequencies 
for the experimental (I) and optimized (II) geometries.
Values are given in meV (cm$^{-1}$).}
\begin{tabular}{lcccc}
 & \multicolumn{2}{c}{MgB$_2$}& \multicolumn{2}{c}{AlB$_2$}\\
Mode & I & II & I  & II \\
\tableline
E$_{1u}$ & 39.9 (322) & 40.5 (327) & 33.0 (266) & 36.6 (295)\\ 
A$_{2u}$ & 48.9 (394) & 50.2 (405) & 48.6 (392) & 52.1 (420)\\ 
E$_{2g}$ & 66.5 (536) & 70.8 (571) & 118.3 (954) & 125.0 (1008)\\  
B$_{1g}$ & 86.3 (696) & 87.0 (702) & 59.9 (483) & 61.3 (494)
\end{tabular}
\end{table}  
\clearpage
\begin{figure}
\caption{Micro-Raman spectra obtained from polycrystalline grains of
MgB$_2$ and AlB$_2$ at room temperature ($\lambda=514.5$ nm)}
\end{figure}
\begin{figure}
\caption{Theoretical phonon dispersion curves along high-symmetry
lines of the hexagonal BZ (notation after \protect\cite{Landolt}) 
and density-of-states
(DOS) of MgB$_2$ and AlB$_2$. Dots represent actually calculated
modes, lines are obtained by Fourier interpolation.}
\end{figure}
\begin{figure}
\caption{Calculated generalized phonon density-of-states of MgB$_2$.
Values for the incoherent scattering cross sections are taken from 
Ref.~\protect\cite{Osborn}.  }
\end{figure}
\begin{figure}
\caption{Calculated Eliashberg functions $\alpha^2 F(\omega)$ 
for MgB$_2$ and AlB$_2$.}
\end{figure}

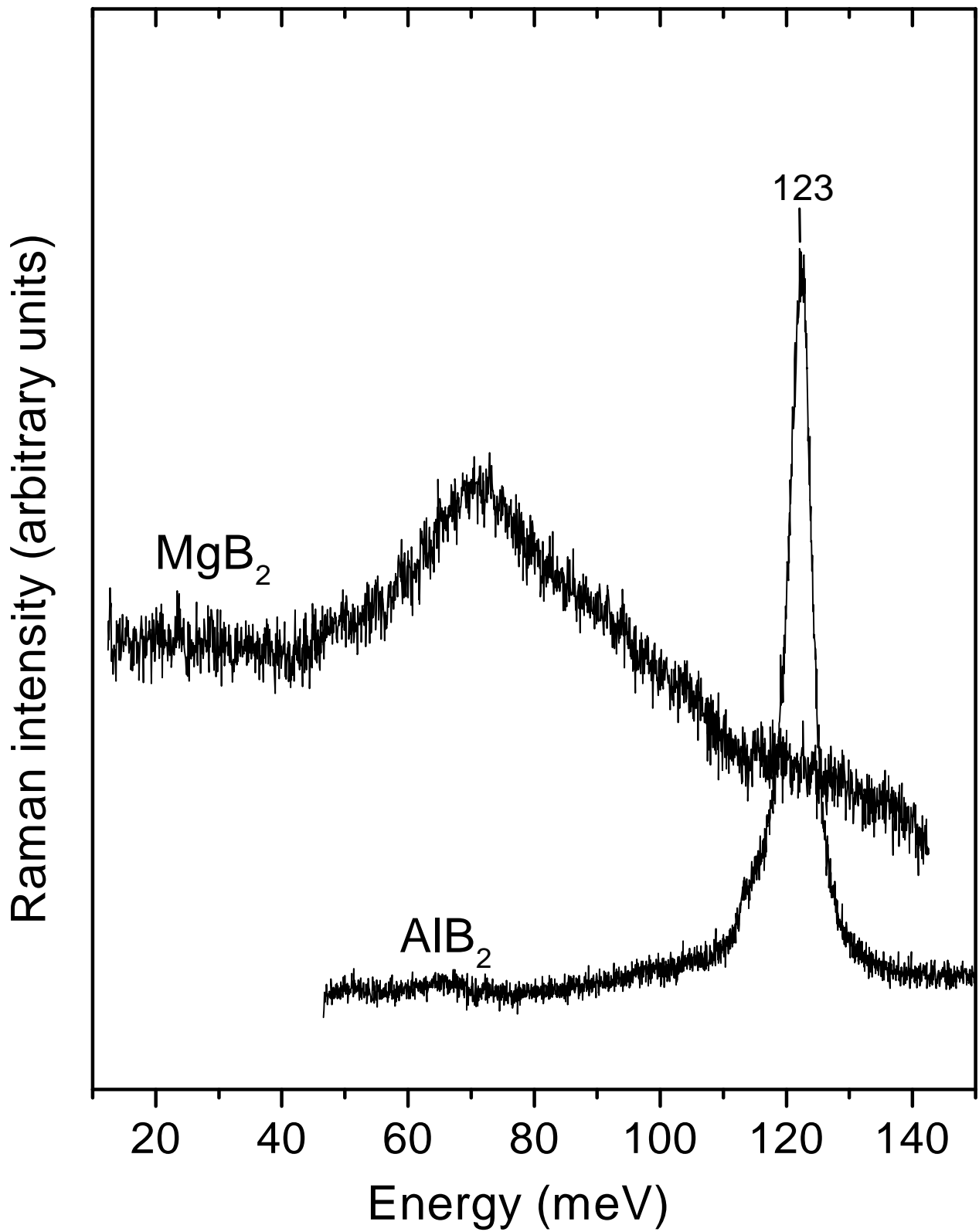
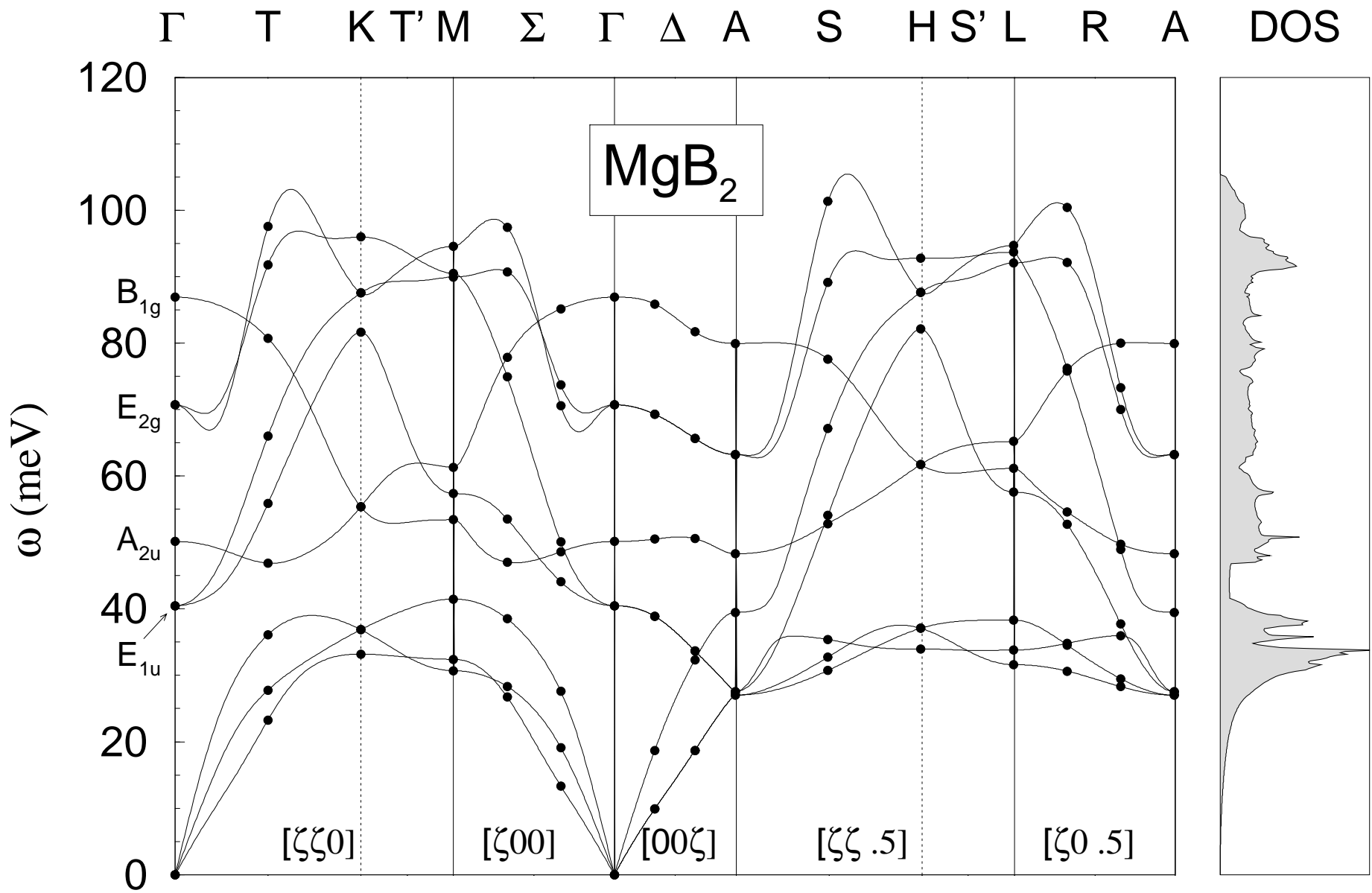
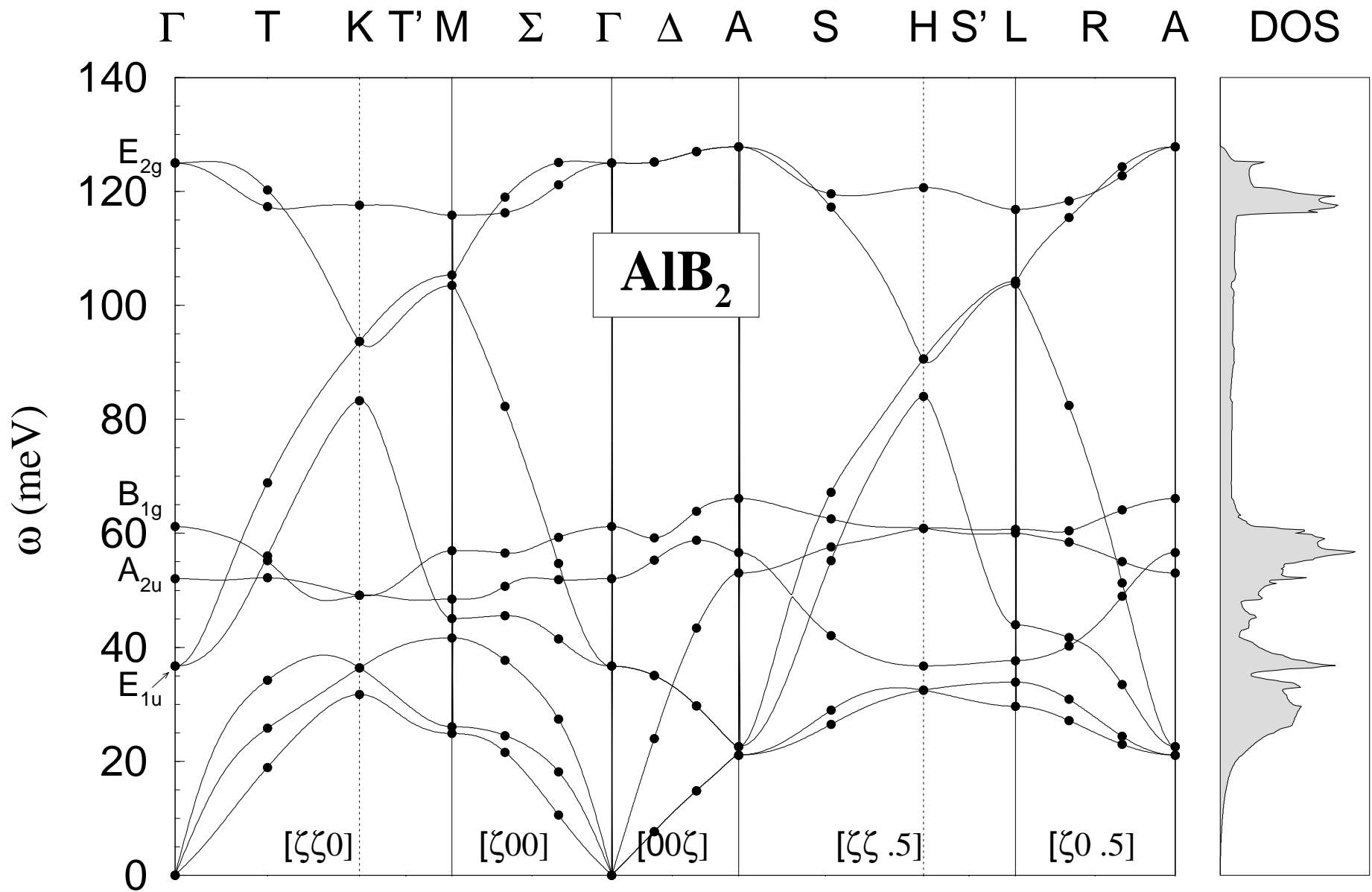
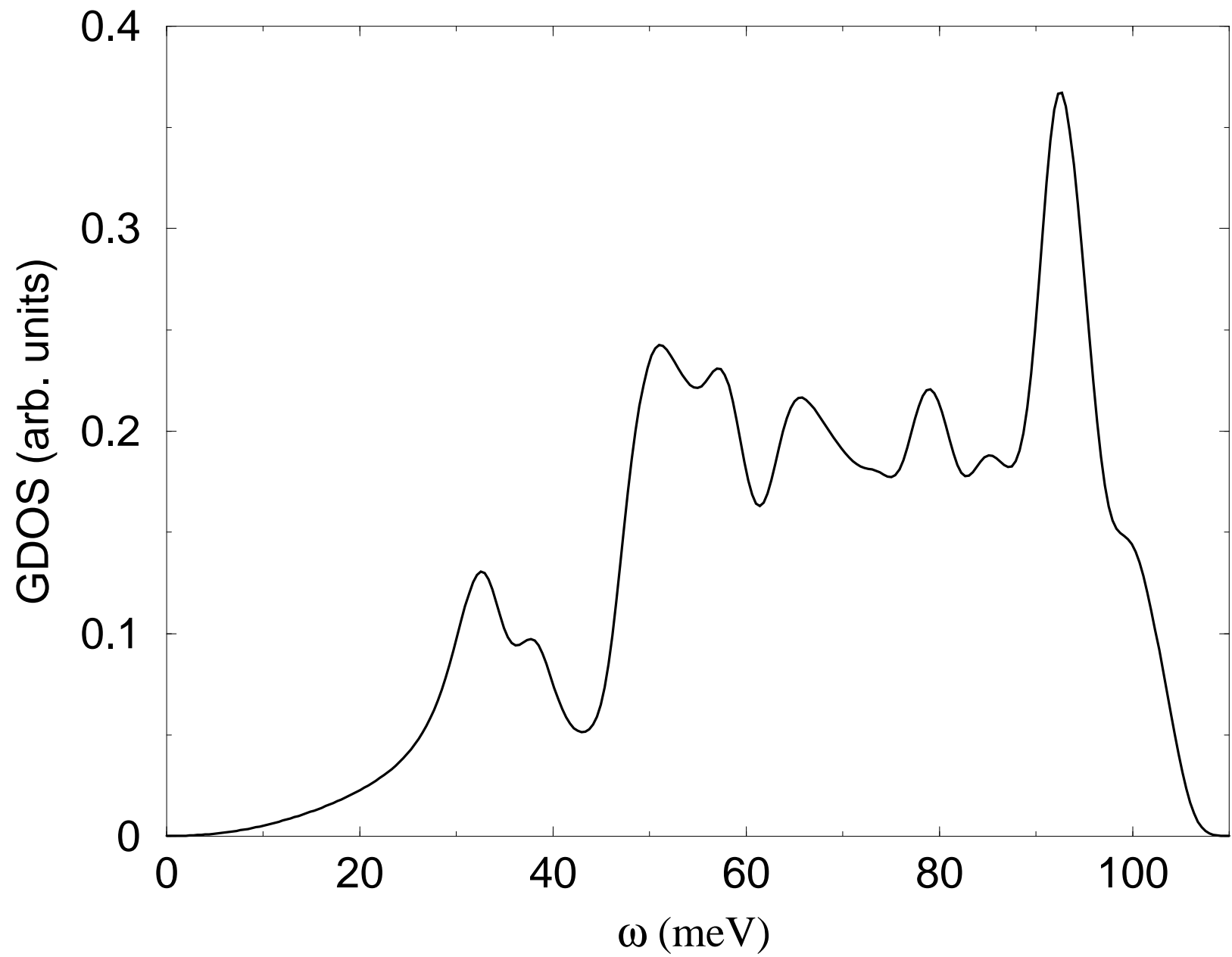
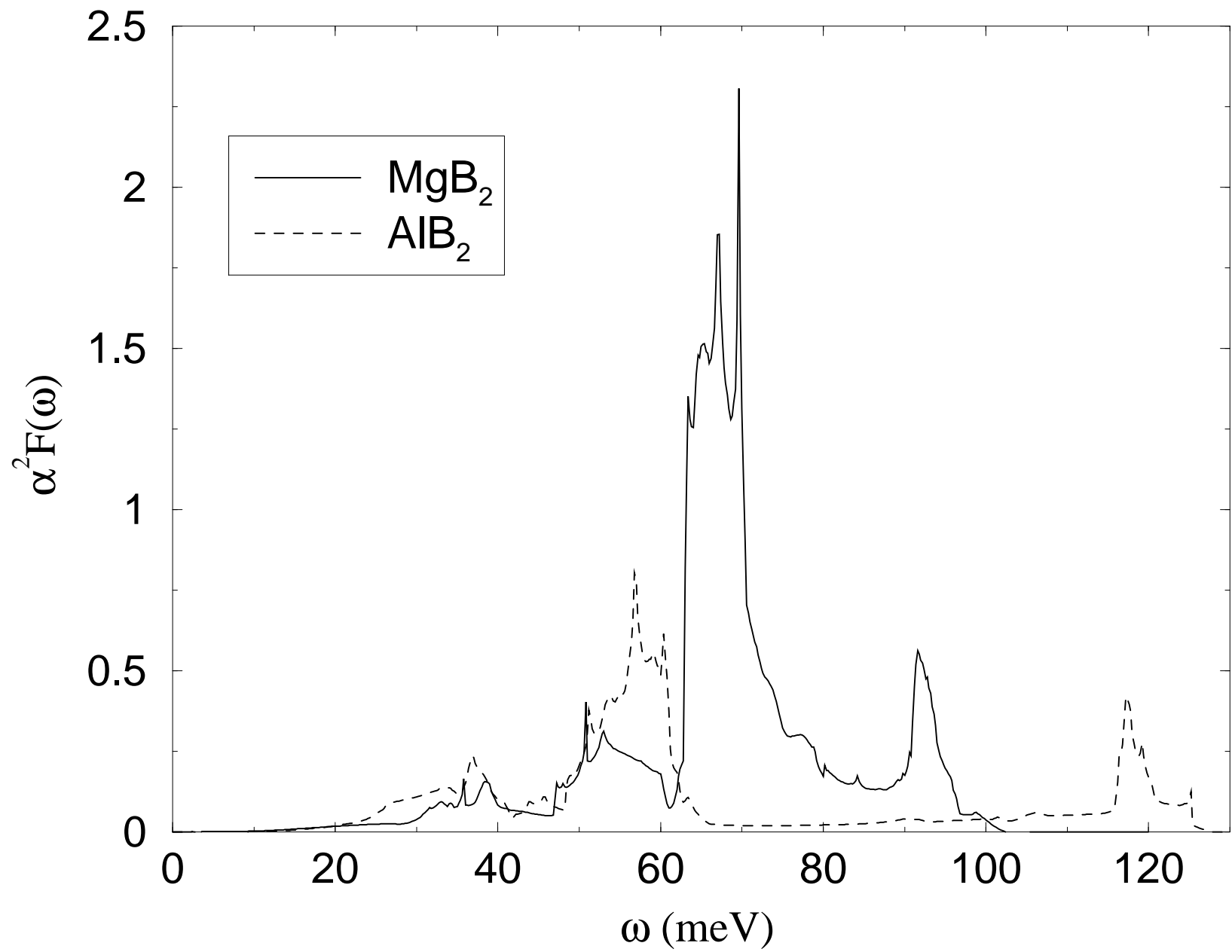
\end{document}